\documentclass[showpacs,twocolumn,floatfix,prl]{revtex4}
\usepackage{graphicx}
\usepackage{color}
\usepackage{psfrag}
\usepackage{epsfig}
\usepackage{bm}
\newcommand{\ket}[1]{| #1 \rangle}

\newcommand{\rb}[1]{\left( #1 \right)}
\newcommand{\mb}[1]{\mathbf{#1}}
\newcommand{\ew}[1]{\langle #1 \rangle}
\newcommand{\beq}{\begin{eqnarray}}
\newcommand{\eeq}{\end{eqnarray}}

\newcommand{\svec}{\mbox{\boldmath$\sigma$}}
\begin{document}
\title{Spin-orbit-driven coherent oscillations in a few-electron quantum
dot}
\author{Stefan Debald}
\affiliation{1.~Institut f\"ur Theoretische Physik,
Universit\"at Hamburg,
Jungiusstr.~9, 20355 Hamburg, Germany}
\author{Clive Emary}
\affiliation{Instituut-Lorentz, Universiteit Leiden, P.O.~Box 9506,
2300 RA Leiden, The Netherlands}
\date{\today}

\begin{abstract}
We propose an experiment to observe coherent oscillations in a
single quantum dot with the oscillations driven by spin--orbit
interaction. This is achieved without spin-polarised leads, and
relies on changing the strength of the spin--orbit coupling via an
applied gate pulse. We derive an effective model of this system
which is formally equivalent to the Jaynes--Cummings model of
quantum optics. For parameters relevant to a InGaAs dot, we calculate a
Rabi frequency of $2\,$GHz.
\end{abstract}

\maketitle
Motivated by the desire for a closer understanding of quantum 
coherence and by the drive to develop novel quantum
computing architecture, a number of breakthrough solid-state
experiments have focused on coherent oscillations --- the back
and forth flopping of that most fundamental of quantum objects, 
the two--level system \cite{nak99,vio02,chi03,hay03}. 
The pioneering work of Nakamura {\it et al.}~with the coherent
superposition of charge states of a Cooper-pair box \cite{nak99} first
demonstrated the possibility of observing such oscillations in a wholly
solid--state device; thus sparking the remarkable progress in qubit
development in super-conducting systems \cite{vio02,chi03}.

The important double quantum dot experiment of
Hayashi and co--workers \cite{hay03} showed that
coherent oscillations could also
be observed in normal semiconductor systems.
It is the purpose of this paper to propose an experiment
in which coherent oscillations are observed in a
{\it single quantum dot} (QD), with these oscillations being driven
by the spin--orbit (SO) interaction.

The SO interaction in semiconductor
heterostructures has its origin in the breaking of inversion
symmetry, and is increasingly coming to be seen as a tool
with which to manipulate electronic states,
see e.g.~\cite{so1}. 
The grandfather of these ideas is the spin--transistor of Datta and
Das \cite{dat90}, in which the SO interaction causes electron
spins to precess as they move through a two-dimensional
electron gas (2DEG). In materials where the structural inversion
asymmetry dominates, e.g.~InGaAs, the SO interaction
can be described by the Rashba Hamiltonian
\cite{byc84}
 \beq
  H_\mathrm{SO} = - \frac{\alpha}{\hbar}
  \left[
    (\mb{p} + \frac{e}{c} \mathbf{A}) \times \svec
  \right]_z.
\label{Hso}
\eeq

In this letter we consider the effects of $H_\mathrm{SO}$
on electrons in a small, few-electron lateral quantum dot.
Although such dots are yet to be realised in materials with
strong SO coupling, there is currently a considerable effort to
develop nanostructures in such materials \cite{zut04}.
Our interest here is not in open or
chaotic QDs \cite{fol01hal01ale01,cre03}, but rather in small dots in
the Coulomb blockade regime.

Such dots have been studied by a number of authors 
\cite{koe04,cha04,sodots}, but our analysis differs in
a crucial respect:  by making an analogy with quantum optics, we
are able to derive an approximate Hamiltonian that captures the
essential physics of the dot.   This model is formally identical
to the Jaynes--Cummings (JC) model \cite{jay63}, first derived in the
context of the atom--light interaction.  Here, the roles of the atomic
pseudo-spin and light field are played by the spin and
orbital angular momentum of the electron respectively.
The system then naturally decomposes into a set of two-level
systems (TLS), any of which may be considered as the
qubit degree of freedom within which coherent
oscillations can occur. These 
oscillations are genuine Rabi oscillations \cite{all75},
with orbital and spin degrees of freedom exchanging excitation.
This ``spin--orbit pendulum'' behaviour has been noted 
in three-dimensional models in nuclear physics \cite{arv94}.

Having elucidated the origin and properties of the TLS, we then
describe an experimental scheme through which the
coherent oscillations can be investigated.  The key
problem here
is that of injecting into, and reading out from,
states which are not eigenstates of the SO
coupled system. In the Hayashi experiment \cite{hay03}, 
this was achieved through the spatial separation of the two dots, which
makes the leads couple to the localised left and right states, rather
than to the bonding and anti-bonding eigenstates. In our single dot
system, the direct analogy of this would be the injection of
spin--polarised electrons. Given the difficulty of interfacing
ferromagnetic leads with semiconductors \cite{zut04}, we
avoid their use by exploiting the fact that
the strength of the SO interaction can be controlled by external gates
\cite{nit97,gru00,kog02}.

Our starting point is the Fock-Darwin theory  of a
single electron in a 2DEG
with parabolic confinement of energy $\hbar \omega_0$ \cite{kow01},
\beq
  H_0 = \frac{(\mb{p} + \frac{e}{c}\mb{A})^2}{2m}
  +\frac{m}{2} \omega_0^2 (x^2 + y^2),
\eeq
where $m$ is the effective mass of the electron.  
Applying a perpendicular magnetic field in the symmetric gauge, 
in second quantised notation we have 
\beq
  H_0 =
  \hbar \widetilde{\omega}(a^\dag_x a_x +a^\dag_y a_y +1)
  +\frac{\hbar \omega_c}{2i}(a_y a_x^\dag - a_x a_y^\dag),
\eeq
with $\omega_c \equiv eB/mc$ and
$\widetilde{\omega}^2 \equiv \omega_0^2 +\omega_c^2/4$.
Introduction of $a_\pm = 2^{-1/2}(a_x \mp i a_y)$ decouples 
the system into eigenmodes of frequency 
$\omega_\pm = \widetilde{\omega}\pm \omega_c/2$.

We now include the Rashba interaction of Eq.~(\ref{Hso}),
for which the coupling strength $\alpha$ is related to the
spin precession length $l_\mathrm{SO} \equiv \hbar^2 / 2 m \alpha$.
With  magnetic length
$l_B \equiv \sqrt{\hbar/m\omega_c}$,
we have
\beq
  H_\mathrm{SO} =  \frac{\alpha}{\tilde{l}}
  \left[
    \gamma_+(a_+\sigma_+ + a_+^\dag \sigma_-)
    - \gamma_-(a_- \sigma_-+a_-^\dag \sigma_+)
  \right],
\eeq
with coefficients
$\gamma_\pm \equiv 1 \pm \frac{1}{2}\rb{\tilde{l}/l_B}^2$
and
$\tilde{l}\equiv\sqrt{\hbar/m \widetilde{\omega}}$.

Adding the Zeeman term, in which we take $g$ to be negative as in InGaAs,
performing a unitary rotation of the spin
such that $\sigma_z\rightarrow -\sigma_z$ and
$\sigma_\pm\rightarrow -\sigma_\mp$,
and rescaling energies by  $\hbar\omega_0$ we arrive at the Hamiltonian
\beq
  H =
  \omega_+ a^\dag_+ a_+
  +\omega_- a^\dag_- a_-
  +\frac{1}{2}E_z  \sigma_z
  ~~~~~~~~~~~ ~~~~~ ~~~~~
  \nonumber \\
  ~~
  +\frac{l_0^2}{2 \tilde{l}\,l_\mathrm{SO}}
    \left[
       \gamma_-(a_- \sigma_+ +a_-^\dag \sigma_-)
      -\gamma_+(a_+\sigma_- + a_+^\dag \sigma_+)
    \right],
  \label{Hfull}
\eeq where $l_0 = \sqrt{\hbar/ m \omega_0}$ is the confinement
length of the dot and $E_z=|g|m/(2 m_\mathrm{e})(l_B/l_0)^2$
is the Zeeman energy
with $m_\mathrm{e}$ the bare mass of the electron.

This single--particle picture is motivated by the good
agreement between Fock-Darwin theory and experiment in the non-SO
case \cite{kow01},
and by studies which have shown that many-body effects in
QDs play only a small role at the magnetic fields we consider here
\cite{koe04,cha04,mer91}. 
\begin{figure}[t]
  \centerline{
    \includegraphics[width=1.0\linewidth,clip=true]{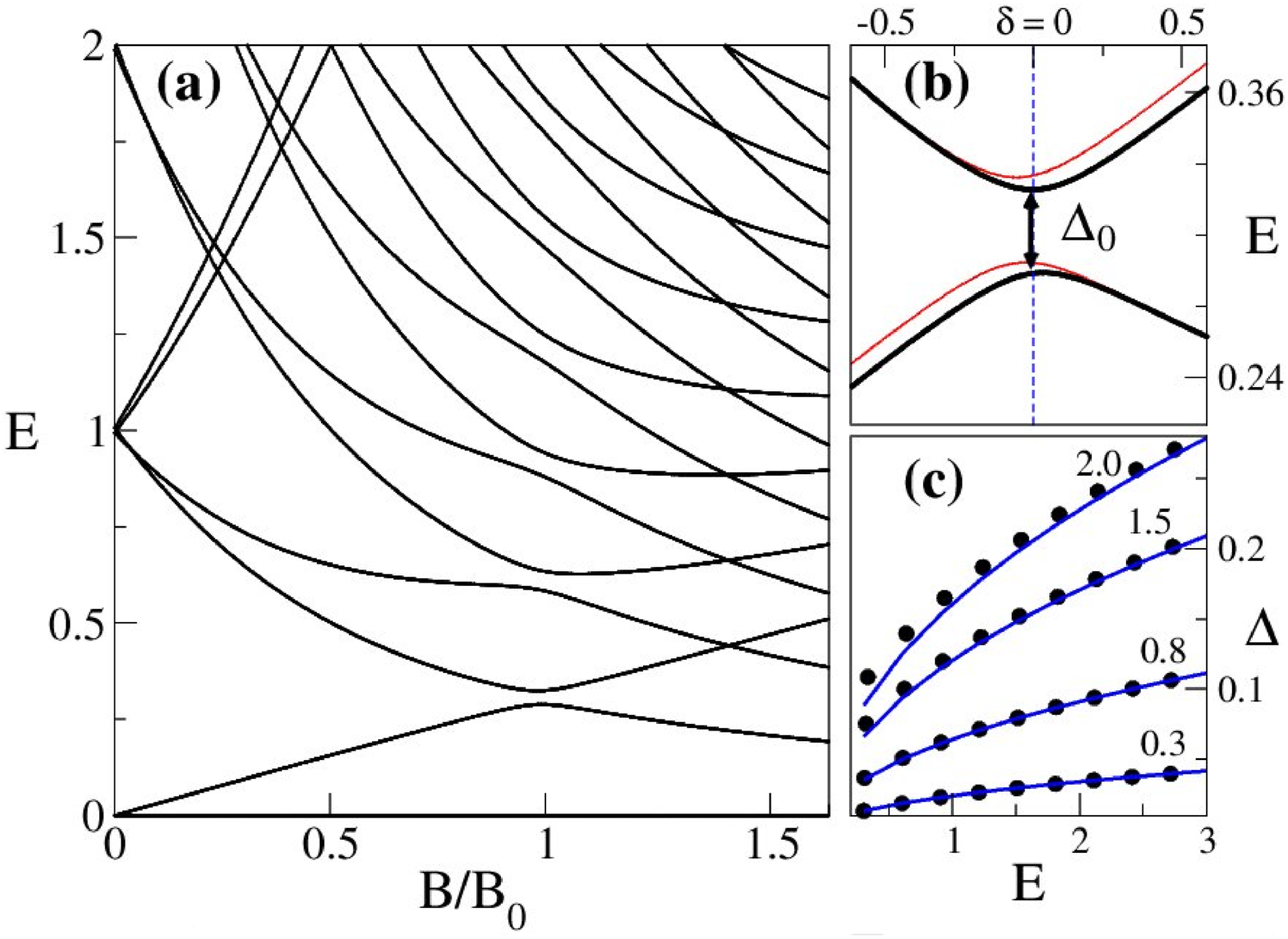}
  }
  \caption{Spectral features of Rashba--coupled quantum dot
  as function of magnetic field.
  The parameters used are typical of InGaAs: $g = -4$,
  $m/m_e = 0.05$ with dot size $l_0=150\,$nm. Resonance occurs at
  $B_0=90\,$mT.
  {\bf (a)} Low--lying excitation spectrum for spin-orbit
  coupling $\alpha=0.8\times 10^{-12}\,$eVm.
  {\bf (b)} Lowest lying anticrossing.  Thick line is JC model showing
  anticrossing width $\Delta_0$ at $\delta=0$, and thin line is
  exact numerical result.
  {\bf (c)} Plot of width $\Delta_n$ against central
  energy of anticrossing with the dot on resonance for different
  $\alpha$
  in the range $0.3 - 2.0 \times 10^{-12}\,$eVm.  The exact numerical
  results
  (circles) show excellent agreement with the square--root behaviour
  predicted by the JC model in this $\alpha$ range.
  \label{spec}
  }
\end{figure}

We now derive an approximate form of this Hamiltonian
by borrowing the observation from
quantum optics that the terms preceded by $\gamma_+$ in
Eq.~(\ref{Hfull}) are counter-rotating, and thus negligible under the rotating--wave approximation  \cite{all75} when the SO coupling is small compared to the confinement.
This decouples the $\omega_+$ mode from the rest of the system, giving
$H =\omega_+ n_+  + H_\mathrm{JC}$
where
\beq
  H_\mathrm{JC}(\alpha)  =  \omega_- a^\dag_- a_-
  + \frac{1}{2}E_z \sigma_z
  +\lambda (a_- \sigma_+ + a_-^\dag \sigma_-),
\label{Hjc}
\eeq
with
$\lambda=l_0^2 \gamma_- /2 \tilde{l}\,l_\mathrm{SO}  $.
This is the well-known Jaynes-Cummings model (JCM) of quantum optics. It
is
completely integrable, and has
ground state $\ket{0,\downarrow}$
with energy $E_\mathrm{G} =- E_z/2$ independent of coupling.
The rest of the JCM Hilbert space decomposes
into two-dimensional subspaces
$\left\{\ket{n,\uparrow},\ket{n+1,\downarrow} ;\quad
n=0,1,\ldots\right\}$.
Diagonalisation in each subspace gives the energies
$E_{\alpha}^{(n,\pm)} = (n+1/2)\, \omega_- \pm  \Delta_n/2$
with detuning $\delta \equiv \omega_- -E_z $ and 
$\Delta_n \equiv \sqrt{\delta^2 + 4 \lambda^2(n+1)}$. 
The eigenstates are
\beq
 \ket{\psi^{(n,\pm)}_\alpha} =
 \cos \theta^{(n,\pm)}_\alpha \ket{n,\uparrow}
  + \sin \theta^{(n,\pm)}_\alpha\ket{n+1,\downarrow},
\label{wfn1}
\eeq
with $\tan\theta^{(n,\pm)}_\alpha = (\delta \pm
\Delta_n)/2\lambda\sqrt{n+1}$.

Figure~\ref{spec}a shows a portion of the excitation spectrum
obtained by exact
numerical diagonalisation for a typical dot in InGaAs. The approximate
$H_\mathrm{JC}$ describes the energy levels of the system to within 10\%
of the typical anticrossing width and 1\% of $\omega_0$. This small
discrepancy is visible in Fig.~\ref{spec}b.
In the following, we are only concerned with the
lowest--lying energy states in the dots.  Without SO interaction,
these states are described by  $n_+=0$ -- indicating that the states
converge to the lowest Landau level in the high--field limit, and
by $n_-$ corresponding to the quantum number of angular momentum.
The SO interaction thus couples two states of adjacent angular
momentum and opposite spin. The detuning $\delta$ uniquely identifies
$\omega_c$ for fixed material parameters and dot size.

Under the assumptions of the constant interaction model \cite{kow01},
the most important prediction of this model for
linear transport is that, with the dot on resonance,
the addition--energy spectrum for the first few electrons
(up to 18 here) is described by a sequence of
well-separated anticrossings, the width of
which increases as $\alpha \sqrt{n+1}$.  This behaviour is shown in
Fig.~\ref{spec}c, and its observation would be confirmation of our
JC model, and would permit a determination of $\alpha$ 
in quantum dots.

\begin{figure}[t]
\begin{center}
\psfrag{(a)}{\bf{(a)}}
\psfrag{(b)}{\bf{(b)}}
\psfrag{(c)}{\bf{(c)}}
\psfrag{(d)}{\bf{(d)}}
\psfrag{pu}{$\psi_{\alpha_1}^+$}
\psfrag{pd}{$\psi_{\alpha_1}^-$}
\psfrag{Gl}{$\Gamma_L$}
\psfrag{Gr}{$\Gamma_R$}
\psfrag{G1}{$\Gamma_1$}
\psfrag{G2}{$\Gamma_2$}
\psfrag{ml}{$\!\mu_L$}
\psfrag{mr}{$\mu_R$}
\psfrag{a11}{$\alpha_1$}
\psfrag{a12}{$\alpha_1$}
\psfrag{a13}{$\alpha_1$}
\psfrag{a2}{$\alpha_2$}
\epsfig{file=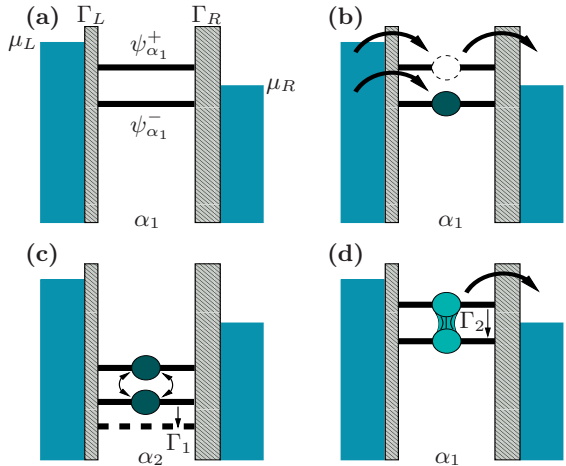, width=0.850\linewidth, 
height = 6cm
}
\end{center}
  \caption{
    Configuration of the dot in the various stages of the cycle.
    {\bf (a)} The positions of the dot levels $\psi_{\alpha_1}^\pm$,
    chemical potentials $\mu_{L,R}$, and the
    tunnelling rates  $\Gamma_L > \Gamma_R$.
    {\bf (b)}  The coupling is initially $\alpha_1$.
    On average, for times $t_i >\Gamma_R^{-1}$ the dot will
    be initialised in state $\psi_{\alpha_1}^-$.
    {\bf (c)} The applied voltage pulse lowers the dot
   levels and non-adiabatically changes the coupling to
   $\alpha_2\neq\alpha_1$, thus inducing Rabi oscillations.
    {\bf (d)} Pulse is switched off after time $t_p$
   and the levels return to their initial places.  Tunnelling to right
   occurs when the electron has oscillated into upper state. Relaxation 
   rates $\Gamma_{1,2}$ are also shown.
  \label{dots}
  }
\end{figure}

We now describe the procedure for observing 
spin-orbit driven Rabi oscillations.
Our proposal is somewhat similar to that of
Nakamura \cite{nak99} with a voltage pulse driving the
system, but with the crucial difference that
the oscillations here are induced, not by a change in the
detuning, but by a change in the SO coupling strength.
We operate in the non-linear transport regime 
and address a
single two-level system by being near resonance and by tuning the
chemical potentials of the leads close to the $n$-th anticrossing.
The SO coupling is set to $\alpha_1$ and the states taking
part in the oscillation are eigenstates of
$H_{\mathrm{JC}}(\alpha_1)$, namely
$\psi_{\alpha_1}^{\pm}$,
which are situated symmetrically around the chemical potential
of the right lead $\mu_R$, see Fig.~\ref{dots}a. 
The temperature is taken smaller than the detuning
$k_{\mathrm{B}} T \ll \delta$ 
to avoid the effects of thermal broadening. Assuming Coulomb blockade
and considering first--order sequential tunnelling only,
electrons can either tunnel from the left lead into the dot via state 
$\psi_{\alpha_1}^{+}$ and subsequently leave to the 
right or, alternatively, tunnel to state $\psi_{\alpha_1}^{-}$ 
blockading  the dot, see
Fig.~\ref{dots}b.  Assuming tunnelling through the left/right
barrier at a constant rate $\Gamma_{L/R}$, 
we set  $\Gamma_L >\Gamma_R$ to assure that the dot
is preferentially filled from the left; thus
maximising the current.
On average then, the dot will be initialised in state
$\psi_{\alpha_1}^{-}$ for times $t_i>\Gamma_R^{-1}$.

Having trapped an electron in this state,
we apply a voltage pulse to the gate.  This has two effects.
Firstly, this change in voltage alters the SO 
coupling to a new value $\alpha_2$.  Since this change is performed
non--adiabatically, the electron remains in the initial eigenstate
$\psi^-_{\alpha_1}$ until Rabi oscillations begin
between this state and $\psi^+_{\alpha_1}$ under the
influence of the new Hamiltonian $H_\mathrm{JC}(\alpha_2)$.
Secondly, the TLS is drawn below both chemical potentials, assuring
that oscillations can occur without tunnelling out of the dot,
see Fig.~\ref{dots}c.

After a time $t_p$, the gate voltage is returned to its
initial value, and the TLS resumes both to its original position and
coupling $\alpha_1$, as in Fig.~\ref{dots}d.
Tunnelling out of the dot can now occur, provided that the electron is
found in the upper state, which happens with a probability given by the
overlap of the oscillating wave function at time $t_p$ with the upper
level,
\beq
  P(t_p) = |\ew{\psi^+_{\alpha_1}|\Psi(t_p)}|^2
  =  |\ew{\psi^+_{\alpha_1}|e^{-i H(\alpha_2)t_p}|\psi^-_{\alpha_1}}|^2.
\label{Ptp}
\eeq

This process is operated as a cycle and the current is measured.  
From probability arguments we see that
$
  I \,\approx \,\,e \Gamma_R P(t_p),
$
where we have used the simplification that
$\Gamma_R^{-1} > t_p,\Gamma_L^{-1} $.
Thus, by sweeping $t_p$ we are able to image the time
evolution of Rabi oscillations,
just as in the previous experiments of Nakamura and
Hayashi.

The singular case of a non-adiabatic change in $\alpha$ from zero to
a finite value 
produces oscillations with the maximum possible amplitude, 
$P_{\mathrm{max}}=1$.
However, in realistic systems only changes between finite values of
$\alpha$ are feasible.
This leads to a reduction in the amplitude, and 
achieving a significant oscillation signal requires a
suitably large change in $\alpha$.
In experiments with 2DEGs, changes in $\alpha$
of a factor of 2 are reported, and in a recent Letter by
Koga {\it et al.}, $\alpha$ was shown to vary in
the range $\approx (0.3 - 1.5) \times 10^{-12}\,$eVm (a factor of 5) in
one InGaAs sample \cite{kog02}. Grundler \cite{gru00} has shown that the
large back--gate voltages usually used to change $\alpha$
can be
drastically reduced by placing the gates closer to the 2DEG.
Thus, it is conceivable that changes in $\alpha$ of a factor between 2
and 5 could be produced with voltages small enough to be pulsed 
with rise times substantially shorter than a typical coherent 
oscillation period.

\begin{figure}[t]
  \centerline{
    \includegraphics[width=0.95\linewidth,clip=true]
{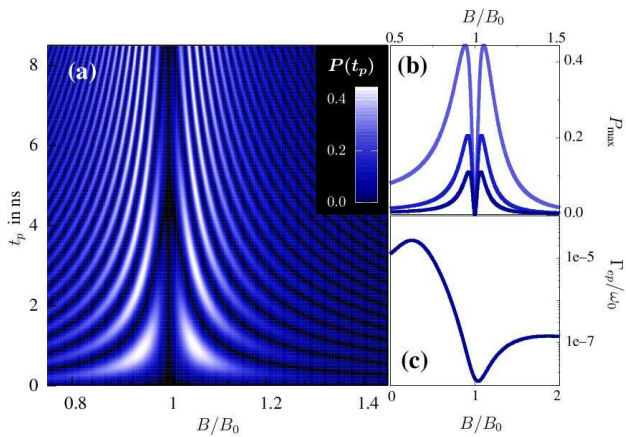}
    }
\caption{Characteristics of the Rabi oscillation.
  \textbf{(a)} Probability $P(t_p)$ of finding electron in upper level
  after time $t_p$ following the non-adiabatic change
  $\alpha_1=1.5\rightarrow  \alpha_2=0.3\times 10^{-12}\,$eVm
  as function of magnetic field.
  \textbf{(b)} Amplitude of oscillation as function of $B/B_0$ for
  changing from $\alpha_1=1.5$, $0.8$, $0.6$ to
  $\alpha_2=0.3\times 10^{-12}\,$eVm (top to bottom).
  \textbf{(c)} Phonon-induced relaxation rate for InAs 
  parameters $\alpha=1.5\times 10^{-12}\,$eVm, 
  $P=3.0\times 10^{-21}\,$J$^2$/m$^2$, $\rho=5.7\times 10^3\,
  $kg/m$^3$, $c=3.8\times 10^{3}\,$m/s. Close to $B_0$ the rate is 
  suppressed to $\Gamma_{ep} < 10^{-7}\omega_0$.
  \label{osc}
}
\end{figure}

In Fig.~\ref{osc}a we plot time--traces of the transition probability
$P(t_p)$ calculated for the first anticrossing
as a function
of magnetic field.  We have used the values
$\alpha_1=1.5 \times 10^{-12}$ and $\alpha_2=0.3 \times 10^{-12}\,$eVm
from the Koga
experiment \cite{kog02}. The amplitude of the oscillations
$P_{\mathrm{max}}$ for different ratios of $\alpha_2/\alpha_1$ is
presented in Fig.~\ref{osc}b, which 
shows a node at $B=B_0$ ($\delta=0$). This is
because, for
$\delta=0$, the eigenstates of JCM are
$2^{-1/2}(\ket{n,\uparrow}\pm\ket{n+1,\downarrow})$ for all $\alpha \ne
0$. Therefore, a finite detuning is required to 
obtain the maximum amplitude, 
which concurs with $\delta>k_{\mathrm{B}}T,\Gamma_R$ to overcome broadening 
effects.
Both the amplitude $P_\mathrm{max}$ and frequency $\Omega$ show
non--trivial dependencies on $\alpha_1$ and $\alpha_2$ as well as
on the magnetic field. This latter behaviour stems from the
parametric dependence on $B$ of all three parameters in $H_\mathrm{JC}$.

For our model parameters with $\alpha_2 / \alpha_1=1/5$
and with the detuning set such that the amplitude is maximised,
we have $P_\mathrm{max} \approx 0.45$ with a Rabi
frequency of $\Omega = 2\,$GHz, which corresponds to a period of about
3\,ns.  This is within accessible range of state--of--the--art
experimental technique. Note that the period can be
extended by using weaker confinement and SO coupling.

For both the observation of coherent oscillations, and the
operation as a qubit, it is essential that the lifetime of state 
$\psi^+_\alpha$ is long.  
This is the case for a pure electronic 
spin in a QD \cite{kha00,zut04}, and we now show that the 
hybridisation of the spin with the orbitals, and the 
ensuing interaction phonons, does not affect this.
We assume a  piezo-electric coupling to acoustic phonons via the potential
$V_{ep}=\lambda_{\bf q} e^{i{\bf q}\cdot{\bf r}} 
(b_{\bf q}+b_{-\bf q}^{\dagger})$, with phonon operators $b_{\bf q}$ and
$|\lambda_{\bf q}|^2=\hbar P/2\rho c q \mathcal{V}$, with
coupling $P$, mass density $\rho$, speed of sound $c$, 
and volume $\mathcal{V}$ \cite{phonons}.  For $n=0$, a Golden Rule calculation yields 
the rate
\beq
  \Gamma_{ep}/\omega_0=\frac{m P}{8\pi(\hbar\omega_s)^2\rho l_0}
  \frac{\sqrt{2}l_0}{\tilde{l}} \sin^2\theta_+\sin^2\theta_-\,\xi^5 I(\xi),
\eeq
with $\omega_s=c/l_0$,  $\xi=2^{-1/2}(\tilde{l}/l_0)(\Delta/\hbar\omega_s)$, 
and $I(\xi)\leq 8/15$. Close to $B_0$, $\xi\ll 1$, and thus the rate
is extremely small $\Gamma_{ep}\approx 10^4\,$s$^{-1}$ 
(Fig.~\ref{osc}c).  Therefore, the robustness of spin qubits
is not significantly weakened by the SO hybridisation.

In general, residual relaxation affects our measurement scheme in two ways. 
During the oscillation (Fig.~\ref{dots}c), the
system may relax to the eigenstate $\psi_{\alpha_2}^-$.  This damps the
oscillation by a factor $\exp(-\Gamma_1 t_p)$ to the constant value 
$I = e \Gamma_R P_{\mathrm{max}}/2$.
Relaxation during the read-out phase (Fig.~\ref{dots}d) simply reduces 
the overall amplitude of the signal by a factor $\exp(-\Gamma_2/\Gamma_R)$.
Clearly then, to observe oscillations, we require  $\Gamma_1 < \Omega$ and
$\Gamma_2 < \Gamma_R$.

In summary, we have outlined a proposal for the observation of
spin-orbit driven coherent oscillations in a single
quantum dot. 
We have derived an approximate model, inspired by quantum optics, that
shows the oscillating degree of freedom to represent a novel, composite
spin-angular momentum qubit.

This work was supported by the EU via TMR/RTN projects,
and the German and Dutch Science Foundations DFG, NWO/FOM.  We are grateful to 
T.~Brandes, C.W.J.~Beenakker and D.~Grundler for discussions, 
and to B.~Kramer for guidance and hospitality in Hamburg.

\end{document}